\documentclass{article}
\usepackage[utf8]{inputenc}

\usepackage{amsmath, amsthm, amssymb}
\usepackage[sort]{natbib}
\usepackage{graphicx}
\usepackage{caption, subcaption}
\usepackage{enumerate}
\usepackage{centernot}
\usepackage{xfrac}

\usepackage{xr}
\newtheorem{definition}{Definition}
\newtheorem{theorem}{Theorem}

\newcommand{\cindep}{\ensuremath{\perp\!\!\!\perp}}
\newcommand{\notcindep}{\ensuremath{\centernot\cindep}}
\newcommand{\given}{\ensuremath{\,|\,}}

\begin{document}

\title{A potential outcomes approach to selection bias}
\author{
 Eben Kenah \\ 
 Biostatistics Division, College of Public Health,\\
 The Ohio State University}
\maketitle

\begin{abstract}
  We propose a novel definition of selection bias in analytic epidemiology using potential outcomes.
  This definition captures selection bias under both the structural approach (where conditioning on selection into the study opens a noncausal path from exposure to disease in a directed acyclic graph) and the traditional definition (where a given measure of association differs between the study sample and the population eligible for inclusion).
  It is nonparametric, and selection bias under this approach can be analyzed using single-world intervention graphs both under and away from the null hypothesis.
  It allows the simultaneous analysis of confounding and selection bias, it explicitly links the selection of study participants to the estimation of causal effects using study data, and it can be adapted to handle selection bias in descriptive epidemiology.
  Through examples, we show that this approach provides a novel perspective on the variety of mechanisms that can generate selection bias and simplifies the analysis of selection bias in matched studies and case-cohort studies.
\end{abstract}

\section{Introduction}
Along with confounding, selection bias is one of the fundamental threats to the validity of epidemiologic research.
Traditionally, selection bias has been defined as a systematic difference (i.e., a difference beyond random variation) between measures of an exposure-disease association in a study sample and the underlying eligible population, where the \emph{study sample} is included in the study and the \emph{eligible population} is eligible for inclusion~\citep{dahabreh2019extending}.
For example, \citet{berkson1946limitations} showed that two diseases can be positively correlated among hospitalized patients even when they are independent in the source population.
This definition can depend on the parameterization of the exposure-disease association~\citep{hernan2017invited}, and it provides little or no guidance about identifying and controlling selection bias in the design and analysis of a study.
These problems resemble those that arise when confounding is defined as a change in an exposure-disease association upon adjustment for a covariate~\citep{miettinen1981confounding, greenland1986identifiability, wickramaratne1987confounding, greenland1999confounding}.

The structural approach of~\citet{hernan2004structural} showed that selection bias occurs when conditioning on selection into the study opens a noncausal path from an exposure $X$ to a disease $D$ in a causal directed acyclic graph~(DAG) \citep{pearl1995causal, greenland1999causal} for the eligible population.
This approach is nonparametric, and its use of DAGs to incorporate background knowledge in the identification and control of selection bias is a practical advantage over the traditional definition.

However, the structural approach only captures selection bias that can occur under the null hypothesis (i.e., no causal path from $X$ to $D$) and no confounding (i.e., no open backdoor path from $X$ to $D$)~\citep{hernan2017invited}.
An example of selection bias that escapes this approach was given in~\citet{greenland1977response}:
In a hypothetical cohort study where right censoring was not associated with exposure, there was no selection bias under the null.
Away from the null, the risk ratio in the full cohort (censored and uncensored) differed from the risk ratio in the observed cohort (uncensored only).

Here, we use potential outcomes~\citep{rubin1974estimating} to propose a novel definition of selection bias in analytic epidemiology, where the goal is to infer the causal effect of a treatment or exposure on the risk of a disease.
The proposed definition allows the simultaneous analysis of confounding and selection bias, and it captures all selection bias under the structural approach as well as the traditional definition.
It is nonparametric, it can be analyzed using single-world intervention graphs~(SWIGs) of~\citet{richardson2013primer, richardson2013single} both under and away from the null, and it explicitly links the selection of study participants to the measures of association that can be estimated using study data.
We show how it can be adapted to handle selection bias in descriptive epidemiology, where the goal is to estimate the joint distribution of disease and covariates (e.g., demographics or exposures).
Through examples, we show how selection bias can be generated by colliders at $X$ and $D$, how it can arise in randomized clinical trials, and how the potential outcomes approach simplifies the analysis of selection bias in matched studies and case-cohort studies.

\subsection{Confounding, exchangeability, and backdoor paths}
Unlike selection bias, confounding has a standard definition in terms of potential outcomes.
Let $X$ be an exposure or treatment, $D$ be a disease outcome, and $D^x$ denote the outcome that occurs if we intervene to set $X = x$.
In the notation of~\citet{dawid1979conditional}, there is no unmeasured confounding when
\begin{equation}
  D^x \cindep X \given C
  \label{eq:confounding}
\end{equation}
(i.e., $D^x$ is conditionally independent of $X$ given $C$) where $C$ is a set of measured nondescendants of $X$~\citep{rosenbaum1983central}. 
The conditional independence in equation~\eqref{eq:confounding} is called \emph{exchangeability}~\citep{greenland1986identifiability}.
This definition of confounding is difficult to use directly as a guide to study design and analysis.

Confounding can also be defined as an open backdoor path from $X$ to $D$ in a causal DAG~\citep{greenland1999causal}, which provides an intuitive way to use background knowledge to identify and control confounding.
The single-world intervention graphs (SWIGs) are transformations of causal DAGs that explicitly represent potential outcomes~\citep{richardson2013primer}.
A SWIG represents the intervention of setting $X = x$ by splitting the node $X$ into two new nodes:
One node represents the realized value of $X$ and inherits all incoming edges from the node $X$.
The other node represents the intervention $X = x$ and inherits all outgoing edges from the node $X$.
The two new nodes are not connected by an edge, and all paths through the node representing the intervention are blocked.
Any node $Y$ that is a descendant of $X$ in the DAG is written $Y^x$ to show that it is a potential outcome that can be observed only in individuals with $X = x$.
If $Y$ is a nondescendant of $X$, then $Y^x = Y$ and can be observed in all individuals.

\begin{figure}
  \centering
  \begin{subfigure}{.45\textwidth}
    \includegraphics[width = \textwidth]{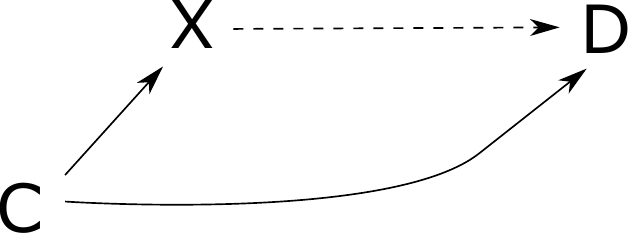}
    \caption{DAG}
  \end{subfigure}
  \hfill
  \begin{subfigure}{.45\textwidth}
    \includegraphics[width = \textwidth]{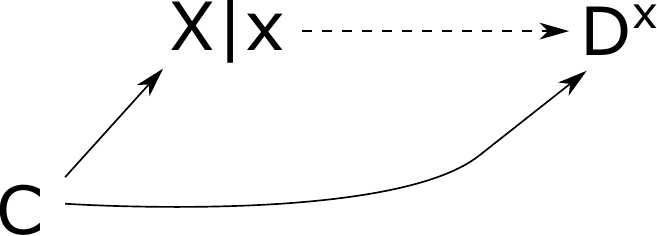}
    \caption{SWIG for setting $X = x$}
  \end{subfigure}
  \caption{
    A causal DAG (left) and SWIG (right) showing no unmeasured confounding given $C$.
    The arrow from $X$ to $D$ is dashed because confounding does not depend on whether $X$ has a causal effect on $D$.
  }
  \label{fig:confounding}
\end{figure}

In a SWIG, the rules of d-separation~\citep{pearl1995causal} can be used to evaluate conditional independence.
This can be used to show that the backdoor path criterion and exchangeability are equivalent~\citep{richardson2013single}.
Figure~\ref{fig:confounding} shows how exchangeability is guaranteed by no open backdoor path from $X$ to $D$ in a simple example:
Conditioning on $C$ blocks the backdoor path from $X$ to $D$ in the DAG, and it d-separates $D^x$ and $X$ in the SWIG.
Therefore, confounding has a nonparametric definition in terms of potential outcomes that can be analyzed using causal graphs both under and away from the null hypothesis.
The potential outcomes approach to selection bias achieves something similar.

\section{Selection bias via potential outcomes}
Let $X$ be an exposure or treatment, $D$ be a disease outcome, and $S$ indicate selection into the study out of a specified eligible population.
The study sample is the subset of the eligible population with $S = 1$.
Assume that $X$ is not a descendant of $D$ in a causal DAG for the eligible population.
We propose the following definition of selection bias in analytic epidemiology:

\begin{definition}[Analytic selection bias]
  There is no unmeasured selection bias for the causal effect of $X$ on $D$ if and only if at least one of the following conditions holds:
  \begin{enumerate}
    \item \textbf{Analytic cohort condition.}
      If we intervene to set exposure $X = x$, selection into the study is conditionally independent of disease outcome given $X$ and $C_1$:
      \begin{equation}
        S^x \cindep D^x \given (X, C_1)
        \label{eq:cohort}
      \end{equation}
      where $C_1$ is a (possibly empty) set of measured nondescendants of $X$ such that exchangeability holds in equation~\eqref{eq:confounding}.
    \item \textbf{Analytic case-control condition.}
      If we intervene to set disease outcome $D = d$, selection into the study is conditionally independent of exposure given $D$, $C_1$, and $C_2$:
      \begin{equation}
        S^d \cindep X \given (D, C_1, C_2)
        \label{eq:casecontrol}
      \end{equation}
      where $C_1$ is defined above and $C_2$ is a (possibly empty) set of measured nondescendants of $D$ such that
      \begin{equation}
        C_2^x \cindep D^x \given (X, C_1).
        \label{eq:c2}
      \end{equation}
      This conditional independence implies that $C_2$ cannot contain variables on a causal path from $X$ to $D$.
  \end{enumerate}
\end{definition}

\begin{figure}
  \centering
  \includegraphics[width = .5\textwidth]{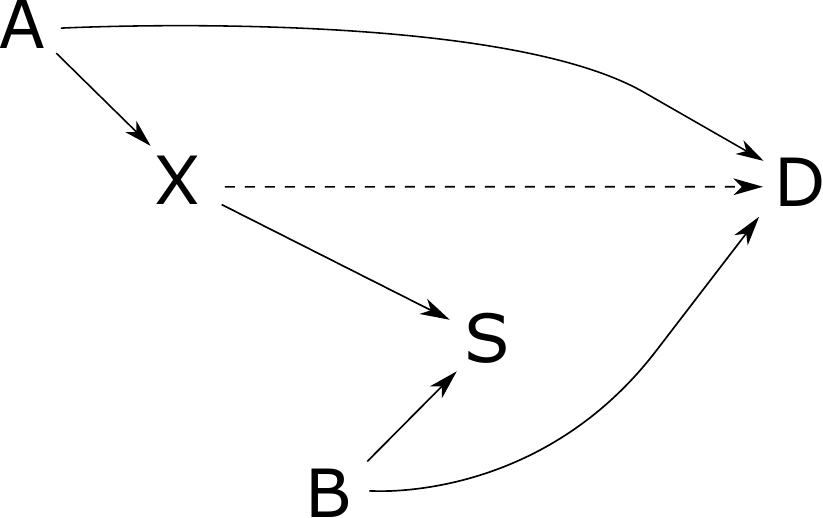} \\
  \medskip
  \begin{subfigure}{.45\textwidth}
    \includegraphics[width = \textwidth]{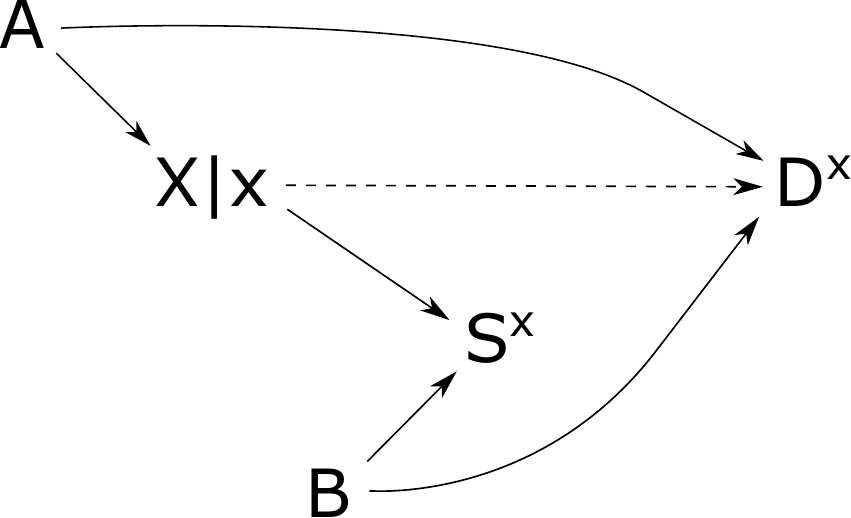}
    \caption{SWIG for setting $X = x$}
  \end{subfigure}
  \hfill
  \begin{subfigure}{.45\textwidth}
    \includegraphics[width = \textwidth]{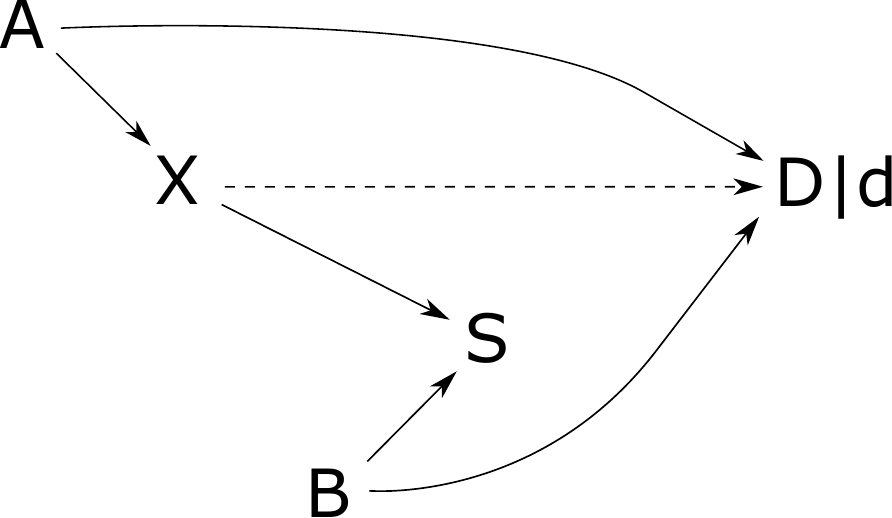}
    \caption{SWIG for setting $D = d$}
  \end{subfigure}
  \caption{
    Causal DAG (top) and SWIGs (bottom) for a the eligible population in a cohort study.
    SWIG~(a) shows that the cohort condition holds given $\{A, B\}$.
    SWIG~(b) shows that the case-control condition fails.
  }
  \label{fig:cohort}
\end{figure}

The analytic cohort condition in equation~\eqref{eq:cohort} is relevant to studies where exposure is a cause of selection.
It is similar to the conditions given in~\citet{daniel2012using} for identifiability of a causal effect with missing data, which is based on the do-operator~\citep{pearl1995causal} rather than potential outcomes.
In the DAG at the top of Figure~\ref{fig:cohort}, we must condition on $A$ to close the backdoor path from $X$ to $D$, so $C_1$ must contain $A$.
In SWIG~(a), we must condition on $B$ to d-separate $S^x$ and $D^x$, so $C_1$ must contain $B$.
The cohort condition holds because $A$ and $B$ are a nondescendants of $X$, exchangeability holds given $C_1 = \{A, B\}$, and $C_1$ d-separates $S^x$ and $D^x$.
The case-control condition fails because of the arrow from $X$ to $S^d = S$ in SWIG~(b).

\begin{figure}
  \centering
  \includegraphics[width = .5\textwidth]{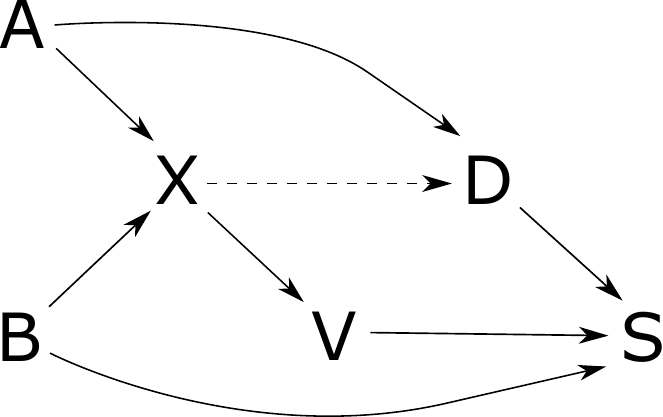} \\
  \medskip
  \begin{subfigure}{.45\textwidth}
    \includegraphics[width = \textwidth]{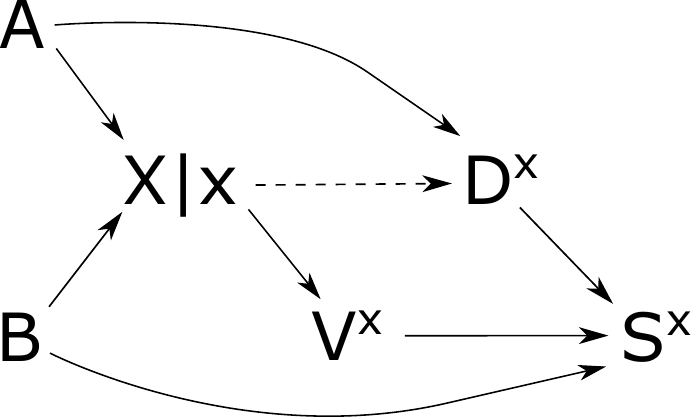}
    \caption{SWIG for setting $X = x$}
  \end{subfigure}
  \hfill
  \begin{subfigure}{.45\textwidth}
    \includegraphics[width = \textwidth]{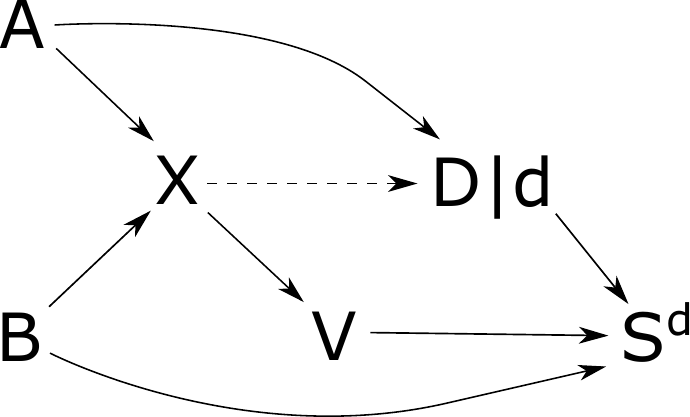}
    \caption{SWIG for setting $D = d$}
  \end{subfigure}
  \caption{
    Causal DAG (top) and SWIGs (bottom) for the eligible population in a case-control study.
    SWIG~(a) shows that the cohort condition fails.
    SWIG~(b) shows that the case-control condition holds given $\{A, B, V\}$.
  }
  \label{fig:casecontrol}
\end{figure}

The analytic case-control condition in equation~\eqref{eq:casecontrol} is relevant to studies where disease outcome is a cause of selection.
It is based on the principle that controls should be individuals who could become cases if they had a disease onset while under observation~\citep{miettinen1985case, wacholder1992selection}.
In Figure~\ref{fig:casecontrol}, the cohort condition fails because of the arrow from $D^x$ to $S^x$ in SWIG~(a).
Because conditioning on $A$ is necessary to block the backdoor path from $X$ to $D$, $C_1$ must contain $A$.
In SWIG~(b), conditioning on $B$ and $V$ is needed to d-separate $X$ and $S^d$.
The variable $V$ cannot be in $C_1$, but it can be in $C_2$ because $V^x \cindep D^x \given (X, A)$.
The variable $B$ can be in $C_1$ because it is a nondescendant of $X$ and exchangeability holds given $\{A, B\}$.
It can also be in $C_2$ because $B^x = B \cindep D^x \given (X, A)$.
Thus, the case-control condition holds given $\{A, B, V\}$ in two ways: $C_1 = \{A\}$ and $C_2 = \{B, V\}$ or $C_1 = \{A, B\}$ and $C_2 = \{V\}$.

Evaluation of the analytic cohort and case-control conditions should consider all stable conditional independencies~\citep{mansournia2015relation} implied by the SWIGs derived from a causal DAG for the eligible population that includes $S$.
The examples above show that neither condition implies the other.
To estimate a conditional causal effect of $X$ on $D$ given a set $V$ of nondescendants of $X$, we must find a $C_1$ that contains all variables in $V$.
The conditions for no unmeasured analytic selection bias are similar to the identifiability conditions for the causal odds ratio in~\citet{bareinboim2012controlling}, but they guarantee the identifiability of a greater variety of causal effect measures.

When there is no unmeasured analytic selection bias, a study has external validity in the sense that adjustment for measured variables is sufficient to generalize a causal effect estimate from the study sample to the eligible population or a subset of it defined by measured variables~\citep{dahabreh2019extending}.
As in~\citet{greenland2011adjustments}, \emph{adjustment for $C$} means a measure of association standardized to a specified joint distribution of the variables in $C$, a set of conditional measures of association within strata of $C$, or a common $C$-conditional measure of association.

\subsection{Selection bias due to a collider at $S$}
\label{sec:colliderS}
\begin{figure}
  \centering
  \includegraphics[width = .4\textwidth]{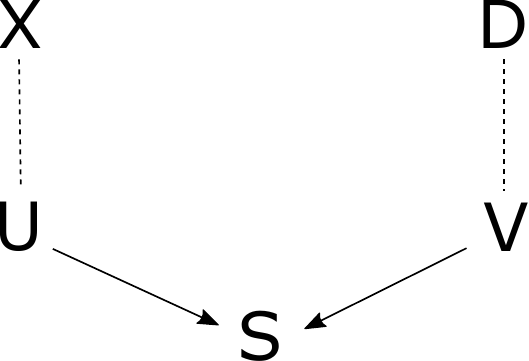}
  \caption{
    Causal DAG showing a variable $S$ as a collider on an open path from $X$ to $D$ in the eligible population.
    Selection into the study is $S$ or a descendant of $S$.
    The dashed lines indicate open paths whose structure is not specified.
  }
  \label{fig:colliderS}
\end{figure}

The structural approach of~\citet{hernan2004structural} identifies selection bias when conditioning on $S$ opens a noncausal path from $X$ to $D$.
This occurs when $S$ is a collider on a path from $X$ to $D$ in a causal DAG or a descendant of a such a collider.
The DAG in Figure~\ref{fig:colliderS} represents a path from $X$ to $D$ on which $S$ is a collider.
As in~\citet{greenland1999causal}, the undirected dashed edges represent open paths whose structure is not specified.
Thus, there is an open path from $X$ to $S$ that ends with an arrow pointing into $S$ and an open path from $S$ to $D$ that starts with an arrow pointing into $S$.

\begin{theorem}
  Selection bias under the structural approach implies analytic selection bias under the potential outcomes approach.
\end{theorem}
\begin{proof}
There is selection bias under the structural approach if and only if at least one path matching the pattern in Figure~\ref{fig:colliderS} exists such that the paths from $X$ to $S$ and from $S$ to $D$ cannot be blocked.
If we condition on $S$ or a descendant of $S$, then:
  \begin{itemize}
    \item If we set $X = x$, the open path from $D$ to $S$ implies that the analytic cohort condition fails in the SWIG for setting $X = x$.
      This holds whether or not $S$ is a descendant of $X$.
    \item If we set $D = d$, the open path from $X$ to $S$ implies that the analytic case-control condition fails in the SWIG for setting $D = d$.
      This holds whether or not $S$ is a descendant of $D$.
  \end{itemize}
  Because the analytic cohort and case-control conditions fail, we have analytic selection bias under the potential outcomes approach.
\end{proof}

In each example from~\citet{hernan2004structural}, the potential outcomes approach and the structural approach reach identical conclusions about the presence and control of selection bias.
As noted by~\citet{hernan2017invited}, all of these examples have no causal path and no open backdoor path from $X$ to $D$.
Selection bias under the structural approach compromises both the internal and external validity of a study:
The causal effect estimate within the study sample is biased, so it cannot generalize to the eligible population.

\subsection{Selection bias in descriptive epidemiology}
The traditional definition of selection bias applies to a measure of association between $X$ and $D$ whether or not it represents a causal effect.
To define selection bias for descriptive epidemiology, we can drop the requirement that exchangeability holds given $C_1$ and remove all restrictions on descendants of $X$ and $D$ so there is no need to distinguish between $C_1$ and $C_2$:

\begin{definition}[Descriptive selection bias]
  There is no unmeasured selection bias for the association between $X$ and $D$ if and only if at least one of the following conditions holds:
  \begin{enumerate}
    \item \textbf{Descriptive cohort condition.}
      Selection into the study is conditionally independent of disease outcome given $X$ and $C$:
      \begin{equation}
        S \cindep D \given (X, C)
      \end{equation}
      where $C$ is a (possibly empty) set of measured covariates.
    \item \textbf{Descriptive case-control condition.}
      Selection into the study is conditionally independent of exposure given $D$ and $C$:
      \begin{equation}
        S \cindep X \given (D, C)
      \end{equation}
      where $C$ is defined above.
    \end{enumerate}
\end{definition}

When at least one of these conditions holds, the conditional association between $X$ and $D$ given $C$ in the study sample matches the same conditional association in the eligible population (up to random variation).
In the cohort study from Figure~\ref{fig:cohort}, conditioning on $B$ controls descriptive selection bias but not analytic selection bias.
The same is true of conditioning on $\{B, V\}$ in the case-control study from Figure~\ref{fig:casecontrol}.
There are two primary differences between the control of descriptive and analytic selection bias:
Analytic selection bias must be controlled using a set of variables sufficient to control confounding, and conditioning on causal descendants of $X$ and $D$ is constrained.

The descriptive cohort and case-control conditions can be assessed on any DAG---causal or not---that represents the joint distribution of $X$, $D$, $S$, and $C$ in the eligible population.
This definition of no unmeasured descriptive selection bias matches the necessary and sufficient conditions given in~\citet{didelez2010graphical} for the $X$-$D$ odds ratio given $C$ to be collapsible over $S$.

\subsection{Selection and estimation}
\label{sec:estimation}
\begin{figure}
  \centering
  \includegraphics[width = .5\textwidth]{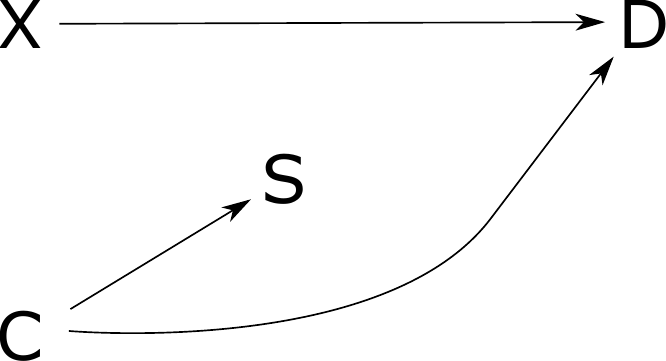} \\
  \medskip
  \begin{subfigure}{.45\textwidth}
    \includegraphics[width = \textwidth]{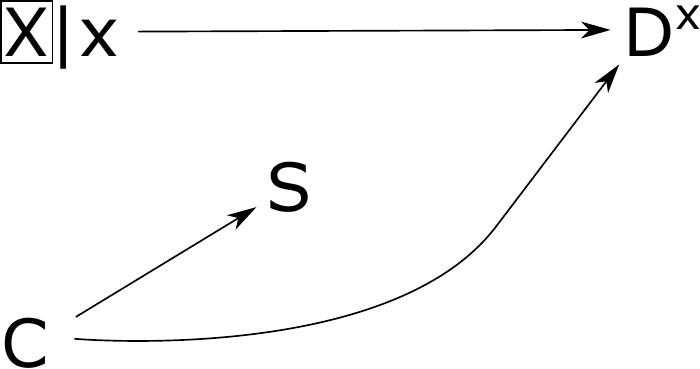}
    \caption{SWIG for setting $X = x$}
  \end{subfigure}
  \hfill
  \begin{subfigure}{.45\textwidth}
    \includegraphics[width = \textwidth]{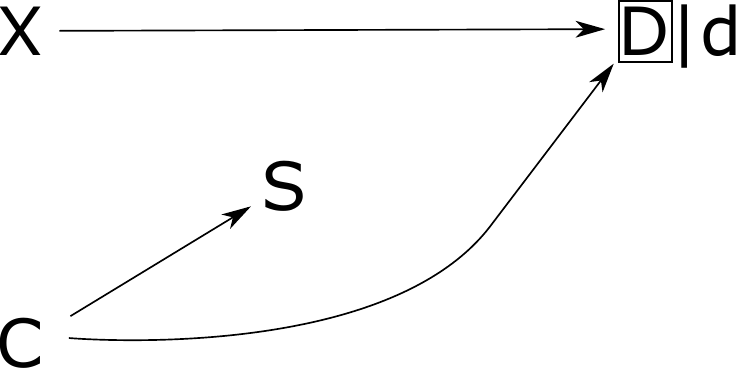}
    \caption{SWIG for setting $D = d$}
  \end{subfigure}
  \caption{
    Causal DAG adapted from~\citet{hernan2017invited} and corresponding SWIGs for the example in~\citet{greenland1977response}.
    The boxes in the SWIGs represent conditioning on $X$ and $D$ when checking the cohort and case-control conditions.
  }
  \label{fig:greenland}
\end{figure}

The selection bias example of~\citet{greenland1977response} was analyzed by~\citet{hernan2017invited} using a DAG similar that in Figure~\ref{fig:greenland}, where $S$ is not a collider or a descendant of a collider.
\citet{howe2016selection} considered a similar causal structure for selection bias caused by loss to follow-up.
In the SWIG for setting $X = x$, the open path $S$-$C$-$D^x$ must be blocked by conditioning on $C$.
In the SWIG for setting $D = d$, the path $S$-$C$-$D$-$X$ is opened by conditioning on the collider at $D$, so it must be closed by conditioning on $C$.
In both SWIGs, the potential outcomes approach correctly identifies selection bias and shows that it can be controlled by adjusting for $C$.

\begin{theorem}
  Selection bias under the traditional definition implies selection bias under the potential outcomes approach.
\end{theorem}
\begin{proof}
  We will prove the equivalent statement that no selection bias under the potential outcomes approach implies no selection bias under the traditional definition.
  If there is no selection bias under the potential outcomes approach, then at least one of the cohort and case-control conditions holds.
  In each case, we will consider both analytic and descriptive selection bias.
  
  When the analytic cohort condition holds and we intervene to set $X = x$, the conditional risk of disease given $C_1 = c_1$ in the eligible population equals the conditional risk given $C_1 = c_1$ and $X = x$ in the study sample:
  \begin{equation}
    \begin{aligned}
      \Pr\bigl(D^x = d \given C_1 = c_1\bigr)
      &= \Pr(D^x = d \given C_1 = c_1, X = x) \\
      &= \Pr\bigl(D^x = d \given C_1 = c_1, X = x, S^x = 1\bigr) \\
      &= \Pr\bigl(D = d \given C_1 = c_1, X = x, S = 1\bigr)
    \end{aligned}
    \label{eq:acohort}
  \end{equation}
  by exchangeability, the cohort condition, and consistency.
  Any measure of causal effect based on conditional risks of disease given $C_1$ in the study sample equals the same measure based on potential outcomes in the eligible population (up to random variation).

  Under the descriptive cohort condition, we have
  \begin{equation}
    \Pr(D = d \given C = c, X = x) = \Pr(D = d \given C = c, X = x, S = 1)
    \label{eq:dcohort}
  \end{equation}
  by the descriptive cohort condition.
  Any measure of association based on conditional risks of disease given $C$ in the study sample equals the same measure in the eligible population (up to random variation).
  
  For the case-control conditions, assume $D$ is binary and that we are comparing two levels of exposure that we call $X = 1$ and $X = 0$ without loss of generality.
  If the analytic case-control condition holds and we intervene to set $X = x$, the conditional odds of disease given $C_1$ is
  \begin{equation}
    \begin{aligned}
      \frac{\Pr(D^1 = 1 \given C_1 = c_1)}{\Pr(D^1 = 0 \given C_1 = c_1)}
      &= \frac{\Pr(D^1 = 1 \given C_1 = c_1, X = x)}{\Pr(D^1 = 0 \given C_1 = c_1, X = x)} \\
      &= \frac{\Pr(D^1 = 1 \given C_1 = c_1, X = x, C_2^x = c_2)}{\Pr(D^1 = 0 \given C_1 = c_1, X = x, C_2^x = c_2)} \\
      &= \frac{\Pr(D = 1 \given C_1 = c_1, X = x, C_2 = c_2)}{\Pr(D = 0 \given C_1 = c_1, X = x, C_2 = c_2)}
    \end{aligned}
    \label{eq:dor}
  \end{equation}
  by exchangeability, the conditional independence in equation~\eqref{eq:c2}, and consistency.
  By Bayes' rule, the final odds equals
  \begin{equation}
    \frac{\Pr(X = x \given C_1 = c_1, C_2 = c_2, D = 1)}{\Pr(X = x \given C_1 = c_1, C_2 = c_2, D = 0)}
    \times \frac{\Pr(D = 1 \given C_1 = c_1, C_2 = c_2)}{\Pr(D = 0 \given C_1 = c_1, C_2 = c_2)}.
  \end{equation}
  Because the second term does not depend on $X$, it cancels out of the conditional causal odds ratio for disease given $C_1$.
  This causal odds ratio equals the conditional odds ratio for exposure given $C_1$ and $C_2$ in the eligible population:
  \begin{equation}
    \begin{aligned}
      &\frac{\sfrac{\Pr(D^1 = 1 \given C_1 = c_1)}{\Pr(D^1 = 0 \given C_1 = c_1)}}
        {\sfrac{\Pr(D^0 = 1 \given C_1 = c_1)}{\Pr(D^0 = 0 \given C_1 = c_1)}} \\
      &\qquad = \frac{\sfrac{\Pr(X = 1 \given C_1 = c_1, C_2 = c_2, D = 1)}{\Pr(X = 0 \given C_1 = c_1, C_2 = c_2, D = 1)}}
        {\sfrac{\Pr(X = 1 \given C_1 = c_1, C_2 = c_2, D = 0)}{\Pr(X = 0 \given C_1 = c_1, C_2 = c_2, D = 0)}}.
    \end{aligned}
  \end{equation}
  Each component of the conditional odds ratio for exposure in the eligible population can be estimated using data from the study sample because
  \begin{equation}
    \begin{aligned}
      &\Pr(X = x \given C_1 = c_1, C_2 = c_2, D = d) \\
      &\qquad = \Pr(X = x \given C_1 = c_1, C_2 = c_2, D = d, S^d = 1) \\
      &\qquad = \Pr(X = x \given C_1 = c_1, C_2 = c_2, D = d, S = 1)
    \end{aligned}
    \label{eq:acasecontrol}
  \end{equation}
  by the case-control condition and consistency.
  Thus, the causal odds ratio for disease given $C_1$ in the eligible population equals the conditional odds ratio for exposure given $C_1$ and $C_2$ in the study sample (up to random variation).

  Under the descriptive case-control condition,
  \begin{equation}
    \begin{aligned}
    &\frac{\sfrac{\Pr(D = 1 \given X = 1, C = c)}{\Pr(D = 0 \given X = 1, C = c)}}
      {\sfrac{\Pr(D = 1 \given X = 0, C = c)}{\Pr(D = 0 \given X = 0, C = c)}} \\
    &\qquad = \frac{\sfrac{\Pr(X = 1 \given D = 1, C = c)}{\Pr(X = 0 \given D = 1, C = c)}}
      {\sfrac{\Pr(X = 1 \given D = 0, C = c)}{\Pr(X = 0 \given D = 0, C = c)}}.
    \end{aligned}
  \end{equation}
  The latter odds ratio can be estimated using the study sample because
  \begin{equation}
    \Pr(X = x \given C = c, D = d) = \Pr(X = x \given C = c, D = d, S = 1)
    \label{eq:dcasecontrol}
  \end{equation}
  by the descriptive case-control condition.
  Thus, the conditional odds ratio for disease given $C$ in the eligible population equals the conditional odds ratio for exposure given $C$ in the study sample (up to random variation).
\end{proof}

The cohort and case-control conditions have different implications for the estimation of causal effects or associations.
The cohort condition allows any measure based on conditional risks of disease to be calculated, including marginal measures based on standardization.
The case-control condition allows only conditional odds ratios to be estimated.
With case-control or case-cohort data, calculating conditional risks of disease requires external information about the eligible population~\citep{miettinen1976estimability}.

\subsection{Selection bias and adjustment}
\label{sec:generalization}
Selection bias under the potential outcomes approach does not imply selection bias under the traditional definition.
Under certain conditions, adjustment can leave a measure of association unchanged~\citep{greenland2011adjustments}.
Adjustment for $C$ will alter a measure of association if there is effect modification by $C$ or noncollapsibility.
When we must condition on $C$ to avoid selection bias, it is important to measure the covariates in $C$ to check these conditions even if adjustment proves unnecessary.

\citet{lu2022toward} proposed a classification of analytic selection bias for the risk difference and risk ratio into forms that occur due to conditioning on a collider (or a descendant of a collider) and forms that occur due to conditioning on an effect measure modifier.
Because both measures are collapsible, these categories probably account for almost all selection bias where adjustment of the risk difference or risk ratio is necessary.
The potential outcomes definition is nonparametric, so it identifies selection bias whenever it is not precluded by the structure of the causal DAG in the eligible population.

\subsection{Analytic versus descriptive selection bias}
The conditions given in~\citet{didelez2010graphical} for collapsibility of an $X$-$D$ odds ratio over $S$ do not ensure that an $X$-$D$ odds ratio that collapses over $S$ is causal~\citep{bareinboim2012controlling}.
This is a special case of the fact that analytic selection bias can occur when there is no descriptive selection bias.
However, no analytic selection bias implies no descriptive selection bias.

\begin{theorem}
  No analytic selection bias implies no descriptive selection bias, but analytic selection bias can occur without descriptive selection bias.
\end{theorem}
\begin{proof}
  Assume the analytic cohort condition holds given a set of variables $C_1$.
  By consistency and equation~\eqref{eq:acohort}, we have:
  \begin{equation}
    \begin{aligned}
      \Pr(D = 1 \given C_1 = c_1, X = x)
      &= \Pr(D^x = 1 \given C_1 = c_1, X = x) \\
      &= \Pr(D = 1 \given C_1 = c_1, X = x, S = 1)
    \end{aligned}
  \end{equation}
  If we let $C = C_1$, this is equivalent to equation~\eqref{eq:dcohort}, which guarantees no descriptive selection bias.

  Now assume the analytic case-control condition holds given $C_1$ and $C_2$.
  If we let $C = C_1 \cup C_2$ (i.e., the union of $C_1$ and $C_2$), then equation~\eqref{eq:acasecontrol} is equivalent to equation~\eqref{eq:dcasecontrol}, which guarantees no descriptive selection bias.

  In the DAG at the top of Figure~\ref{fig:cohort}, the descriptive cohort condition holds given $B$.
  However, the analytic cohort condition fails because conditioning on $A$ is required to block the backdoor path from $X$ to $D$.
  The analytic case-control condition also fails, so there is analytic selection bias given $B$.
  In the DAG at the top of Figure~\ref{fig:casecontrol}, the descriptive case-control condition holds given $\{B, V\}$.
  However, the analytic case-control condition fails because conditioning on $A$ is required to block the backdoor path from $X$ to $D$.
  The analytic cohort condition also fails, so there is analytic selection bias given $\{B, V\}$.
\end{proof}

\section{Applications to study design and analysis}
\label{sec:applications}
The potential outcomes approach to selection bias provides a novel perspective on the variety of mechanisms that can generate selection bias, and it correctly handles cases where adjustment is necessary for generalization to the eligible population even though the unadjusted causal effect or association is valid in the study sample.
It simplifies the analysis of matched studies and case-cohort studies by eliminating the need to consider the cancellation of associations along different paths in a DAG~\citep{mansournia2013matched}.
The Supplemental Digital Content contains examples implemented in \texttt{R}~\citep{Rmanual}.

\subsection{Selection bias due to a collider at $X$}
\label{sec:colliderX}
\begin{figure}
  \centering
  \includegraphics[width = .5\textwidth]{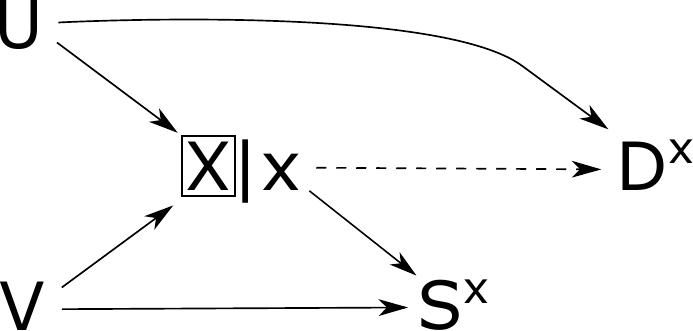}
  \caption{
    SWIG showing selection bias caused by a collider at $X$.
    The box around $X$ represents conditioning on $X$ to check the cohort condition.
    The edge from $X$ to $D$ is dashed because the example does not depend on a causal path from $X$ to $D$.
  }
  \label{fig:colliderX}
\end{figure}

Measures of association based on the risk of disease condition on $X$ when calculating risks within exposure groups.
When there is an open backdoor path from $X$ to $D$, this conditioning can cause descriptive selection bias.
Figure~\ref{fig:colliderX} shows an example.
The descriptive case-control condition fails because of the arrow from $X$ to $S$ in the underlying causal DAG.
In the descriptive cohort condition, conditioning on the collider at $X$ opens the path $S^x$-$V$-$X$-$U$-$D^x$.
However, $S^x \cindep D^x \given (X, C)$ where $C = \{U\}$, $C = \{V\}$, or $C = \{U, V\}$.
The descriptive cohort condition holds given any of these sets.

The analytic cohort condition holds only given $\{U\}$ or $\{U, V\}$ because conditioning on $U$ is necessary to close the backdoor path $X$-$U$-$D$.
Conditioning on $\{V\}$ controls descriptive selection bias but not confounding, so the conditional association between $X$ and $D$ given $V$ is the same in the study sample and the eligible population but differs systematically from the conditional causal effect of $X$ on $D$ given $V$ in the eligible population.
This form of selection bias requires a backdoor path from $X$ to $D$, so it falls outside the range of selection bias considered by~\citet{hernan2004structural}.

\subsection{Selection bias due to a collider at $D$}
\label{sec:colliderD}
\begin{figure}
  \centering
  \includegraphics[width = .5\textwidth]{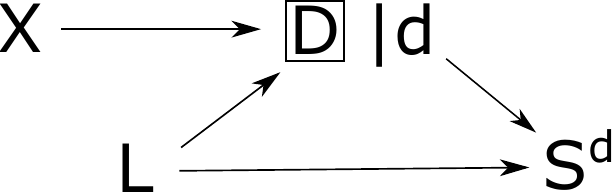}
  \caption{
    SWIG showing selection bias caused by a collider at $D$.
    The box around $D$ represents conditioning on $D$ to check the case-control condition.
    This example also works if the causal path from $X$ to $D$ is replaced by an open backdoor path.
  }
  \label{fig:colliderD}
\end{figure}

Measures of association in case-control studies condition on $D$ when calculating exposure prevalences among cases and controls.
Figure~\ref{fig:colliderD} shows selection bias caused by a collider at $D$ in a case-control study.
The analytic and descriptive cohort conditions fail because of the arrow from $D$ to $S$ in the underlying causal DAG.
In the analytic and descriptive case-control conditions, conditioning on the collider at $D$ opens the path $S^d$-$L$-$D$-$X$.
This path can be blocked by conditioning on $L$, so the descriptive case-control condition holds given $\{L\}$.
The variable $L$ can be in $C_1$ because it is not a descendant of $X$.
It cannot be in $C_2$ because $L^x = L \notcindep D^x \given X$.
Thus, the analytic case-control condition holds given $C_1 = \{L\}$ and $C_2 = \varnothing$ (the empty set).
This form of selection bias cannot occur if there is no causal path and no open backdoor path from $X$ to $D$, so it also falls outside the range of selection bias considered by~\citet{hernan2004structural}.

\subsection{Randomized clinical trials}
\begin{figure}
  \centering
  \includegraphics[width = .5\textwidth]{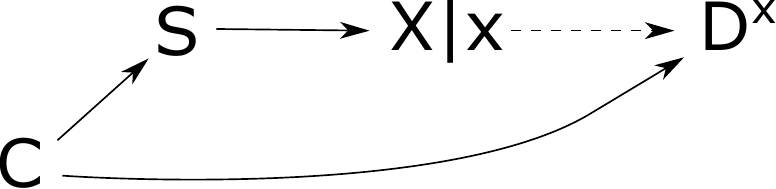}
  \caption{
    SWIG for setting $X = x$ in the eligible population for a randomized clinical trial, where $S$ precedes assignment of $X$ and there is a common cause of $S$ and $D$.
    When selection is related to risk factors for disease, there is a backdoor path from $X$ to $D$ in the eligible population even if $X$ is randomized.
  }
  \label{fig:clinical}
\end{figure}

Figure~\ref{fig:clinical} shows a randomized clinical trial in which there is a common cause $C$ of disease and selection (e.g., selection into the study is based on risk factors for disease).
The arrow from $S$ to $X$ exists because selection into the trial affects the probability of treatment, which might be near zero outside the study sample.
Although randomization of $X$ prevents a backdoor path from $X$ to $D$ in causal DAG for the study sample, there is a backdoor path $X$-$S$-$C$-$D$ in the causal DAG for the eligible population.

The analytic and descriptive cohort conditions both hold given $C$.
In the eligible population, randomization of $X$ ensures that:
\begin{itemize}
  \item \label{item:Sblock} All backdoor paths from $X$ to $D$ are blocked by conditioning on $S$, so the effect of $X$ on $D$ is unconfounded within the study sample.
  \item \label{item:Cdist} Because $C \cindep X \given S$, all treatment groups have the same distribution of $C$.
    Thus, a crude measure of association between $X$ and $D$ is implicitly standardized to the distribution of $C$ in the study sample.
\end{itemize}
Therefore, a randomized trial provides a valid estimate of the causal effect of $X$ on $D$ in the study sample without adjustment for $C$.
However, generalization to the eligible population can require adjustment for $C$ if there is effect modification or noncollapsibility~\citep{greenland2011adjustments}.

\subsection{Matched cohort studies}
\begin{figure}
  \centering
  \begin{subfigure}{.45\textwidth}
    \includegraphics[width = \textwidth]{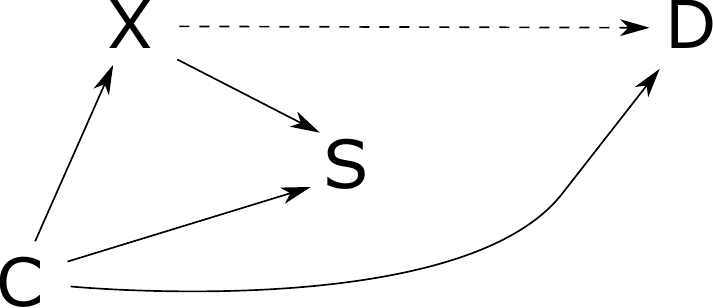}
    \caption{DAG}
  \end{subfigure}
  \hfill
  \begin{subfigure}{.45\textwidth}
    \includegraphics[width = \textwidth]{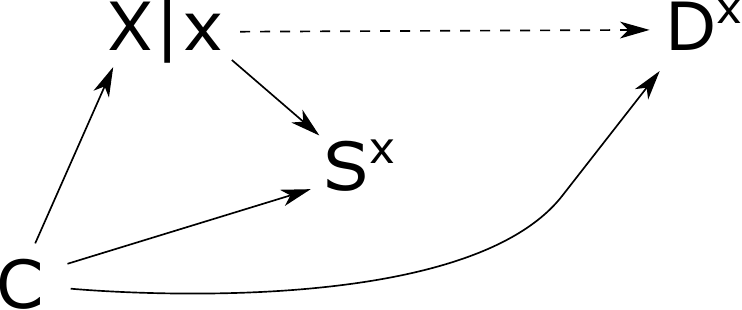}
    \caption{SWIG for setting $X = x$}
  \end{subfigure}
  \caption{
    Causal DAG (left) and SWIG for setting $X = x$ (right) in the eligible population for a cohort study matched on a confounder $C$. 
    Because of matching, all exposure groups have the same distribution of $C$ in the study sample even though $X$ and $C$ are d-connected in the DAG.
  }
  \label{fig:mcohort}
\end{figure}

Figure~\ref{fig:mcohort} shows a cohort study matched on a confounder $C$. 
The analytic cohort condition holds given $C$ because $S^x \cindep D^x \given (X, C)$, so the descriptive cohort condition also holds.
If matching ensures that the distribution of $C$ is the same in all exposure groups in the study sample, then $C \cindep X \given S$ even though $C$ and $X$ are d-connected given $S$.
In this case, a crude measure of association based on disease risks is implicitly standardized to a common distribution of $C$, so no adjustment for $C$ is needed to estimate the marginal causal effect of $X$ within the study sample.
If matching ensures that distribution of $C$ in each exposure group in the study sample matches the distribution of $C$ among the exposed in the eligible population, then this unadjusted estimate corresponds to the average treatment effect among the treated (ATET) in the eligible population.
Generalization requires adjustment for $C$ if there is effect modification or noncollapsibility and a marginal causal effect is being estimated for a different distribution of $C$ than in the study sample~\citep{greenland2011adjustments}.

\subsection{Matched case-control studies}
\begin{figure}
  \centering
  \begin{subfigure}{.45\textwidth}
    \includegraphics[width = \textwidth]{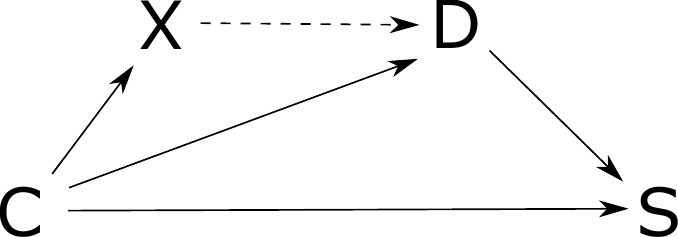}
    \caption{DAG}
  \end{subfigure}
  \hfill
  \begin{subfigure}{.45\textwidth}
    \includegraphics[width = \textwidth]{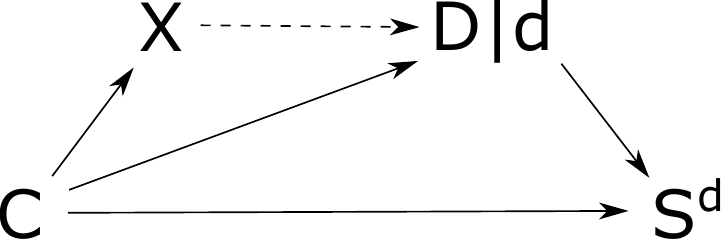}
    \caption{SWIG for setting $D = d$}
  \end{subfigure}
  \caption{
    Causal DAG (left) and SWIG for setting $D = d$ (right) in the eligible population for a case-control study matched on a confounder $C$.
    Because of matching, the case and control groups---not the exposure groups---have the same distribution of $C$.
  }
  \label{fig:mcasecontrol}
\end{figure}

Figure~\ref{fig:mcasecontrol} shows a case-control study matched on a confounder $C$.
Matching ensures that the distribution of $C$ is equal in the case and control groups, not in the exposure groups, so the crude odds ratio in the study sample does not have a causal interpretation.
The $C$-conditional odds ratios have a causal interpretation, and they are identical (up to random variation) in the study sample and the eligible population.

\subsection{Case-cohort studies} 
\begin{figure}
  \centering
  \includegraphics[width = .5\textwidth]{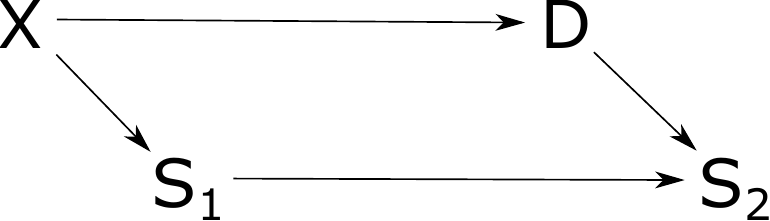} \\
  \medskip
  \begin{subfigure}{.45\textwidth}
    \includegraphics[width = \textwidth]{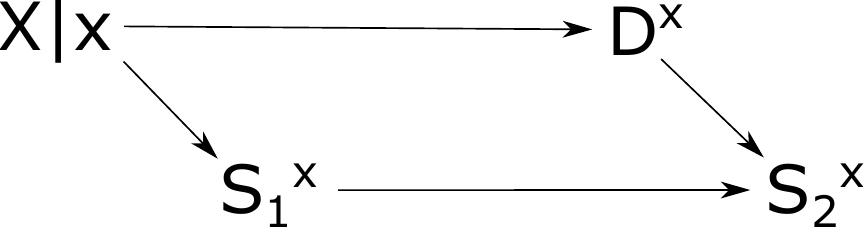}
    \caption{SWIG for setting $X = x$}
  \end{subfigure}
  \hfill
  \begin{subfigure}{.45\textwidth}
    \includegraphics[width = \textwidth]{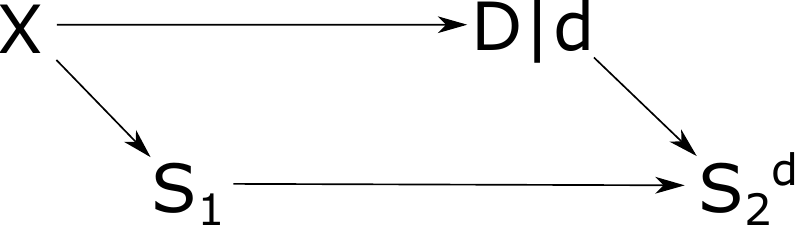}
    \caption{SWIG for setting $D = d$}
  \end{subfigure}
  \caption{
    DAG (top) and SWIGs (bottom) for the eligible population in a case-cohort study. 
    $S_1$ indicates selection into the underlying cohort, and $S_2$ indicates selection into the subcohort or becoming a case (or both). 
  }
  \label{fig:casecohort}
\end{figure}

Figure~\ref{fig:casecohort} shows a causal DAG and corresponding SWIGs for a case-cohort study.
$S_1$ indicates selection into the underlying cohort, and $S_2$ indicates selection into the subcohort or becoming a case (or both).
For selection into the cohort, we have $S_1^x \cindep D^x \given X$ so the analytic cohort condition holds.
For selection into the study sample, the cohort condition fails because of the edge from $D$ to $S_2$, but the analytic case-control condition holds with $C_1 = \varnothing$ and $C_2 = \{S_1\}$.
Because the cohort condition does not hold for $S_2$, the exposure odds ratio comparing cases and controls must be used to estimate the causal effect of exposure on disease.
The need to condition on $S_1$ to control selection bias for $S_2$ implies that cases from outside the cohort must be excluded if the cohort was selected based on exposure.

\section{Discussion}
The potential outcomes approach to selection bias is nonparametric, captures all selection bias under the structural approach of~\citet{hernan2004structural} as well as the traditional definition, and can be adapted to both analytic and descriptive epidemiology.
It is an important practical application of SWIGs, and it provides a unified analysis of confounding and selection bias in analytic epidemiology.
We hope to extend this approach to studies with time-dependent confounding and complex censoring patterns~\citep{robins1986new, robins1992g, robins2000marginal}.

We have assumed throughout that our DAGs completely represent causal relationships in the eligible population.
Causal relationships in a different population might not be represented by the same DAGs.
Thus, no selection bias does not guarantee that an estimated causal effect or association can be generalized to a population containing individuals who were not eligible for the study.
It guarantees generalizability but not transportability~\citep{dahabreh2019extending}, which is a strong argument for the inclusive recruitment of study participants in epidemiology.

\section*{Supplemental Digital Content}
\begin{description}
  % \item Supplemental Digital Content~1: Appendix that uses the approach of Section~\ref{sec:generalization} to analyze the need for adjustment in examples from the text (pdf file).
  \item Supplemental Digital Content: Implementations in \texttt{R}~\citep{Rmanual} of examples from Figures 2-10 (text file).
\end{description}

\section*{Acknowledgements}
I would like to thank Miguel Hern\'{a}n, Forrest Crawford, Sander Greenland, and Patrick Schnell for their useful comments.
This work was supported by National Institute of Allergy and Infectious Diseases (NIAID) grants R01 AI116770 and U01 AI169375 and National Institute of General Medical Sciences (NIGMS) grant U54 GM111274.
The content is solely the responsibility of the author and does not represent the official views or policies of NIAID, NIGMS, or the National Institutes of Health.

\bibliographystyle{chicago}
\bibliography{SSB}

\end{document}